\documentclass[lettersize,journal]{IEEEtran}

\usepackage{enumitem}
\usepackage{afterpage}

\usepackage[nolist]{acronym}
\newacro{SME}[SME]{subject matter expert}
\newacro{US}[U.S.]{United States}

\acused{US}
\acused{SME}

\begin{acronym}
    \acro{AAM}{advanced air mobility}
    \acro{ACC}{adaptive cruise control}
    \acro{ADAS}{advanced driver assistance system}
    \acro{AEB}{automatic emergency braking}
    \acro{AI}{artificial intelligence}
    \acro{ATG}{automatic test generation}
    \acro{AV}{autonomous vehicle}\acro{CA}{California}
    \acro{CADMV}[CA DMV]{California Department of Motor Vehicles}
    \acro{CARLA}{CARs Learning to Act}
    \acro{CE}{Cross Entropy}
    \acro{CP}{correctness property}
    \acrodefplural{CP}{correctness properties}
    \acro{CPS}{cyberphysical system}
    \acro{DL}{deep learning}
    \acro{DNN}{deep neural network}
    \acro{DOD}[DoD]{Department of Defense}
    \acro{DTW}{dynamic time warping}
    \acro{EDA}{electronic design automation}
    \acro{FM}{formal methods}
    \acro{FOL}{first order logic}
    \acro{FVV}[FV\&V]{formal verification and validation}
    \acro{GAN}{Generative Adversarial Network}
    \acro{GPU}{graphics processing unit}
    \acro{HOL}{higher-order logic}
    \acro{HVT}{high-value target}
    \acro{IED}{improvised explosive device}
    \acro{IoT}{internet of things}
    \acro{ISR}{intelligence, surveillance, and reconnaissance}
    \acro{KL}{Kullback-Leibler}
    \acro{LKA}{lane keeping assist}
    \acro{LSTM}{long short-term memory}
    \acro{LTL}{linear-time propositional temporal logic}
    \acro{LVV}[LV\&V]{lightweight verification and validation}
    \acro{ML}{machine learning}
    \acro{MLCP}{machine-learned correctness property}
    \acrodefplural{MLCP}{machine-learned correctness properties}
    \acro{MTL}{metric temporal logic}
    \acro{MTS}{multivariate time series}
    \acro{NASA}{National Aeronautics and Space Administration}
    \acro{NL}{natural language}
    \acro{PDF}{probability distribution function}
    \acro{PED}{processing, exploitation, and dissemination}
    \acro{RD}[R\&D]{research and development}
    \acro{RDF}{random decision forest}
    \acro{RM}{runtime monitoring}
    \acro{RNN}{recurrent neural network}
    \acro{RV}{random variable}
    \acro{SAE}{Society of Automotive Engineers}
    \acro{SITL}{software-in-the-loop}
    \acro{SME}{subject matter expert}
    \acro{STL}{signal temporal logic}
    \acro{SUT}{system under test}
    \acro{SVM}{support vector machine}
    \acro{TEVV}{test and evaluation, verification and validation}
    \acro{TL}{temporal logic}
    \acro{TTC}{time-to-collision}
    \acro{UAS}{unmanned aerial system}
    \acro{VM}{virtual machines}
    \acro{VV}[V\&V]{verification and validation}
    \acro{USN}{\ac{US} Navy}
    \acro{USG}{\ac{US} Government}
\end{acronym}

\usepackage{booktabs}
\usepackage{tabularx}
\usepackage{multirow}
\usepackage{makecell} 
\setcellgapes{5pt}

\usepackage{comment}

\ifCLASSOPTIONcompsoc
  \usepackage[nocompress]{cite}
\else
  \usepackage{cite}
\fi

\ifCLASSINFOpdf
  \usepackage[pdftex]{graphicx}
  \graphicspath{../images/}
\else
\fi

\usepackage{amsmath}
\usepackage{amssymb}
\DeclareMathOperator*{\argmax}{argmax}
\DeclareMathOperator*{\argmin}{argmin}
\usepackage{algorithmic}
\usepackage{algorithm}

\usepackage{array}

\ifCLASSOPTIONcompsoc
  \usepackage[caption=false,font=footnotesize,labelfont=sf,textfont=sf]{subfig}
\else
  \usepackage[caption=false,font=footnotesize]{subfig}
\fi

\usepackage{flushend}

\begin{document}

\noindent{This work has been submitted to the IEEE for possible publication. Copyright may be transferred without notice, after which this version may no longer be accessible.}
\newpage

\title{Discovering Decision Manifolds to\\Assure Trusted Autonomous Systems}

\author{Matthew~Litton,
        Doron~Drusinsky, and
        James~Bret~Michael,~\IEEEmembership{Fellow,~IEEE}
\IEEEcompsocitemizethanks{\IEEEcompsocthanksitem M. Litton, J. Michael, and D. Drusinsky
are with with the Department of Computer Science at the Naval Postgraduate School,
Monterey, CA, 93943.\protect\\
E-mail: \{matthew.litton, ddrusins, bmichael\}@nps.edu
\IEEEcompsocthanksitem This research is sponsored by the U.S. Department of the Navy.}
}


\maketitle

\begin{abstract}
Developing and fielding complex systems requires proof that they are reliably correct with respect to their design and operating requirements.  Especially for autonomous systems which exhibit unanticipated emergent behavior, fully enumerating the range of possible correct and incorrect behaviors is intractable.  Therefore, we propose an optimization-based search technique for generating high-quality, high-variance, and non-trivial data which captures the range of correct and incorrect responses a system could exhibit.  This manifold between desired and undesired behavior provides a more detailed understanding of system reliability than traditional testing or Monte Carlo simulations.  After discovering data points along the manifold, we apply machine learning techniques to quantify the decision manifold's underlying mathematical function.  Such models serve as correctness properties which can be utilized to enable both verification during development and testing, as well as continuous assurance during operation, even amidst system adaptations and dynamic operating environments.  This method can be applied in combination with a simulator in order to provide evidence of dependability to system designers and users, with the ultimate aim of establishing trust in the deployment of complex systems.  In this proof-of-concept, we apply our method to a software-in-the-loop evaluation of an autonomous vehicle.      
\end{abstract}


\section{Introduction}
\acresetall
\IEEEPARstart{P}{roviding} evidence that a system behaves correctly with respect to its design and operating requirements necessitates an \textit{oracle} that can immediately and unambiguously determine if a behavior is correct or incorrect.  For relatively simple systems such as traffic lights, such oracles can be easily encoded in natural language (e.g., if a traffic light is yellow, its next state must be red) or more formal representations like propositional logic (e.g., $CURRENT\_STATE_{YELLOW} \implies NEXT\_STATE_{RED}$).  For such systems, all possible states can be enumerated, and then the system can be tested in its operational environment to guarantee that it performs correctly for a set of test cases that provides complete coverage of all possible system responses.

\begin{figure}[H] 
\centering
\includegraphics[width=0.7\columnwidth]{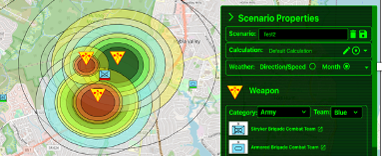}
\caption[]{In other work, we applied our approach to the Defense Threat Reduction Agency's (DTRA) Mission Impact of Nuclear Effects Software (MINES) to discover a decision manifold for military nuclear survivability.  This technique is applicable to any complex system, including autonomous systems, with a realistic simulation capability.}
\label{fig:mines} 
\end{figure}

Autonomous systems pose significant difficulties to this method.  First, the full range of system responses cannot be known, both because many autonomous systems operate in dynamic environments which cannot be fully modeled, and because such systems often use \ac{ML} algorithms to enable continuous adaptation.  Second, testing systems in their operational environment to ensure they behave correctly is not always effective.  Relatively mature systems, such as many \acp{AV}, often exhibit critical failures very rarely, making it unlikely that testing alone will discover those instances.  Other systems, such as nuclear survivability systems, cannot be tested in an operational environment due to the dangers they pose to humans and the environment (see Fig.~\ref{fig:mines}).  To address this challenge, we propose an approach where a system's behavior can be enumerated early in design and testing to discover both correct and incorrect behaviors, including rare events that are unlikely to be observed until well after deployment.  Then, such behaviors are synthesized into a mathematical function representing a \ac{CP} which can be executed on future behaviors, determining if a system's response was correct, even for previously unobserved behaviors.  Then, we propose a similar technique to go beyond monitoring system behavior to actually predicting and assuring correct behavior.  By demonstrating that potential failures can be detected in advance, we provide a means to mitigate the potential for death, injury, damage to property, and damage to the environment that autonomous systems pose.

The core contribution of this paper is the description and implementation of an algorithm (\textit{HybridPair} \ac{CE}) which discovers decision manifolds to verify and assure correct system behavior.  Additionally, we describe our \ac{ML} approach to mathematically quantifying the manifold for use as \acp{CP}.  Then, we show how such \ac{ML} models can be utilized for verifying, validating, and assuring autonomous system behavior to provide evidence of the system's dependability and enable their trusted deployment.

The paper is organized as follows.  In the rest of this section, we clarify terms and definitions related to \acp{CP}, as well as propose a new term: \textit{\acp{MLCP}}.  In Section~\ref{sec:related_work} we discuss related work.  Section~\ref{sec:background} provides background related to the \ac{CE} method and Section~\ref{sec:method} discusses our modification to the algorithm.  Sections~\ref{sec:experiment} and \ref{sec:results_analysis} detail the experiment and its results, and Sections~\ref{sec:discussion_of_results} and \ref{sec:future_work} discuss the implications of the results and avenues for future work.  Finally, Section~\ref{sec:conclusion} concludes the discussion, highlighting the importance of this research.

\subsection{Formal and Practical Properties: Terms and Definitions}\label{subsec:terms}
Several closely related terms are often used in the \ac{VV} literature: \textit{correctness properties}, \textit{formal specifications}, \textit{natural language requirements}, and \textit{assertions}.  As a contribution of our research, we add a fourth term: \textit{machine-learned correctness properties}.  

\begin{figure}[h] 
\centering
\includegraphics[width=0.7\columnwidth]{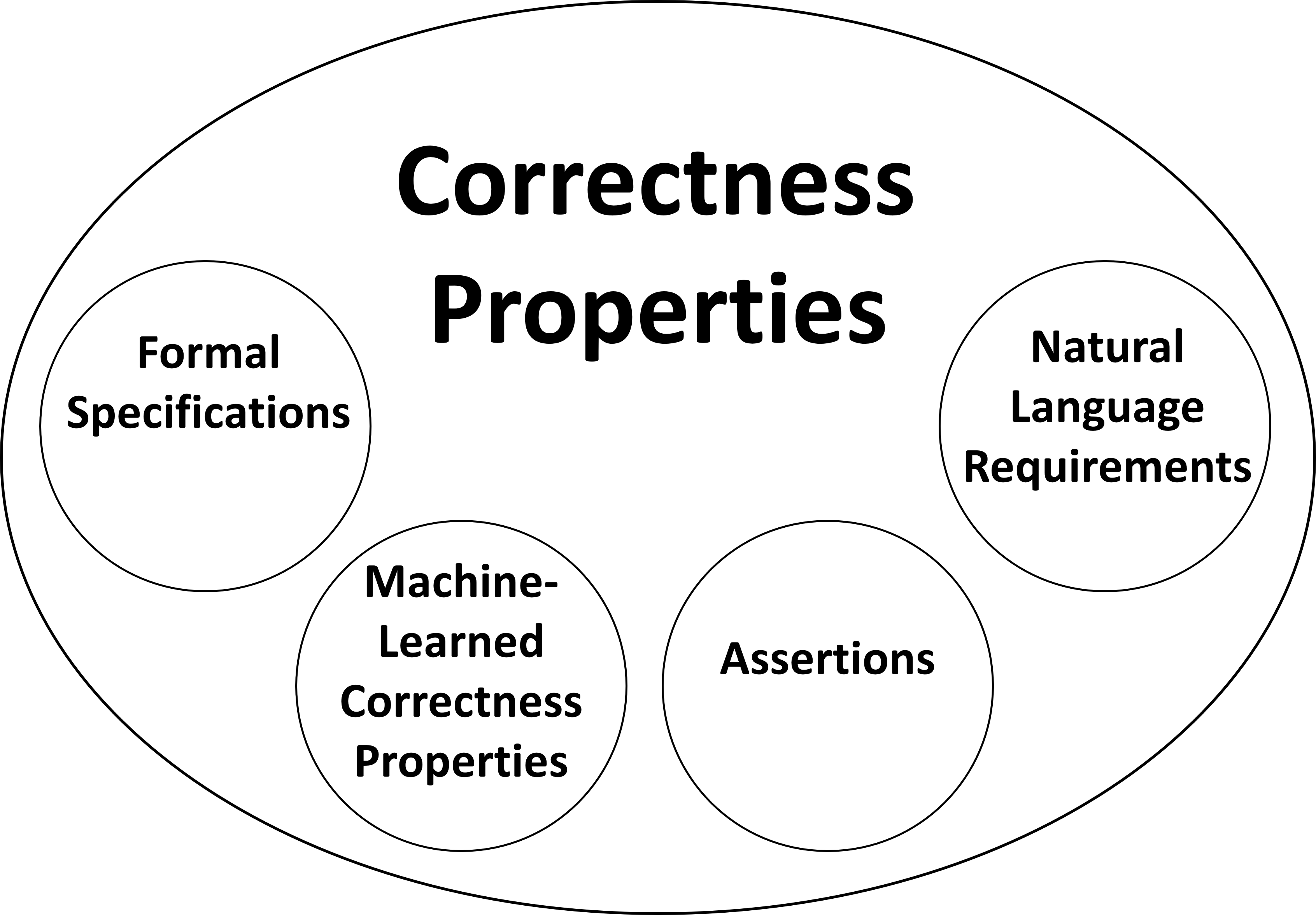}
\caption[]{The relationship between correctness properties and its various sub-types.}
\label{fig:CPs}
\end{figure}

As depicted in Fig.~\ref{fig:CPs}, \textit{correctness properties} is an umbrella term which refers to descriptions of acceptable or unacceptable system behavior.  Often, these requirements are first written in natural language, but ambiguous, context-sensitive languages (e.g., unconstrained English sentences) are not amenable to automated verification.  Therefore, experts manually translate such requirements into an unambiguous, usually machine-readable form.  Historically, the term \textit{specification} has dominated the literature because much of the academic research has focused on human specified properties of interest, using dedicated \textit{formal specification languages}.  \textit{Assertions} typically refer specifically to requirements that are embedded in the source code (e.g., the C library \texttt{<assert.h>}).  

\subsubsection{Machine-Learned Correctness Properties}
We propose an alternative to logic-based \acp{CP} in the form of \acp{MLCP}.  \Acp{MLCP} are \ac{ML} models which are trained to determine whether a system is behaving correctly or incorrectly with respect to some requirement.  They seek the same goal as formally specified properties or embedded assertions which is: to accurately represent correct system behavior so that a system's compliance with its requirements can be verified or monitored.  This idea follows from the ideas presented in our earlier position papers (\cite{litton_2022,litton_waam_2022}), but this work actually describes the construction and evaluation of \acp{MLCP} in detail, and we apply this approach to \acp{AV} to demonstrate their usefulness.  Using \ac{ML} classifiers which are trained using actual system data to discover undesired behavior during testing, and monitor system behavior during operation, addresses weaknesses associated with logic-based \acp{CP}.  First, developing \acp{CP} by training \ac{ML} models brings the task of developing robust \acp{CP} from the realm of Ph.D.-level experts into the domain of system developers and engineers.  Secondly, rather than translating human-specified \acp{CP} into executable code for embedding in the implementation software, \acp{MLCP} automatically learn a model that represents the underlying requirement, and such models are already executable.  Third, the expressive power of \acp{MLCP} extends to the entirety of the operational domain of the \ac{SUT}.  Because \acp{MLCP} use actual system behavior with the environmental configuration, any behavior or interaction that the system can experience, and any configuration of the surrounding environment, can be used to train the model.  

\section{Related Work}\label{sec:related_work}
Two of the areas of research that are related to the research on \acp{MLCP} are: search-based techniques for scenario generation (Section~\ref{subsec:SBT}), and anomaly and incident detection for \acp{CPS} (Section~\ref{subsec:anomaly}).

\subsection{Automated Generation of Scenarios for Cyberphysical Systems}\label{subsec:SBT}
Because the entire state space of most autonomous \acp{CPS} cannot be explored prior to deployment, it is of particular interest to developers (including testers) to identify situations under which a system is most likely to violate its \acp{CP} (e.g., falsification).  These are widely known as "critical events."  The use of falsification to identify such scenarios has been widely explored in the literature.  In \cite{cameron_2015}, the authors model an artificial pancreas controller for Type-1 diabetics, and demonstrate the use of S-Taliro, a falsification engine, for finding violations of \ac{MTL}-specified \acp{CP}.  In nearly all cases, these methods are still largely applied to isolated systems (i.e., without modeling external actors or environmental factors) with finite state spaces.  Other researchers have opted for statistical methods to identify critical events.  Researchers used Monte Carlo sampling of stochastic models of driving behavior to evaluate \acp{AV} in the presence of human drivers, but they discovered the occurrence of unsafe events was very low \cite{zhao_2017}.  In addition, generating such scenarios in naturalistic driving conditions is expensive and slow.  To accelerate the search for events of interest, researchers have used importance sampling methods which can generate a new probability distribution with a higher likelihood of critical events.  The authors in \cite{zhao_2017} use the \ac{CE} method to find an importance sampling distribution that identifies unsafe lane-change behavior by human drivers to accelerate the evaluation of \ac{AV} software stacks.  Similarly, O'Kelly et al. developed a custom simulation framework that utilizes adaptive importance sampling methods to significantly accelerate the evaluation of deep-learning based \acp{AV} compared to open-road testing \cite{okelly_2018}.  Likewise, Ding et al. explored using deep generative models as an alternative to importance sampling techniques for generating safety-critical scenarios \cite{ding_2020}.

\subsection{Anomaly and Incident Detection for Cyberphysical Systems}\label{subsec:anomaly}
Multiple researchers have proposed methods of detecting incorrect behavior in \acp{CPS} through anomaly detection.  Chen et al. \cite{chen_2018} used supervised learning approaches to learn and monitor invariants of a water treatment testbed.  They collect execution traces from the system under normal operating conditions, and then inject faults into the system's underlying software, simulating malicious cyber attacks, to collect execution traces under abnormal operating conditions.  They then use \acp{SVM} to learn a model of correct behavior and monitor the system during runtime.  However, they find that they must generate over 500 mutants (i.e., faults injected into the software) in order to accurately learn invariants; verifying that the mutations accurately reflect realistic malicious cyber activity, as well as generating and simulating the mutated code can be time consuming for certain applications.  In \cite{afzal_2022}, the authors use unsupervised clustering techniques to automatically learn oracles that can determine if a \ac{CPS} is behaving correctly.  They apply a multi-step procedure to cluster telemetry logs, making the assumption that incorrect behavior is likely to be rare, while correct behavior occurs frequently.  Xiao et al. \cite{xiao_2023} note that predicting imminent incidents, including collisions, is a key enabler for the reliable deployment of \acp{AV}.  In their survey, they note that in addition to rule-based and probabilistic methods for predicting hazardous events, \ac{ML}-based methods can allow greater generalization and scalability when learning causal factors for hazardous events \cite{xiao_2023}.  

\section{The CE Method}\label{sec:background}
Section~\ref{subsec:cross_entropy} provides an overview of the \ac{CE} method and is meant to facilitate the reader's understanding of Section~\ref{sec:method} where we describe how we use the \ac{CE} method to generate data for training and testing \ac{ML} models which serve as \acp{MLCP}.    

\subsection{Cross-Entropy Method}\label{subsec:cross_entropy}
The \ac{CE} method derives its name from the cross-entropy (a.k.a., \ac{KL}) distance which is a fundamental concept in information theory.  It was originally developed for estimating the probability of rare events based on variance minimization, and has been adapted for solving combinatorial optimization problems \cite{rubinstein_2004}.  Certain optimization and estimation problems are static and can be solved analytically; many estimation problems are \textit{noisy}, meaning the objective function is unknown and must be estimated.  The dependability analysis of complex systems, including \acp{CPS}, is one such problem.

At its most basic level, the Cross-Entropy Method is an iteration of the two steps of Algorithm~\ref{alg:basic_ce}.

\begin{figure}[!ht]
\begin{algorithm}[H]
\caption{One Iteration of the Basic Cross-Entropy Algorithm}
\label{alg:basic_ce}
\begin{algorithmic}[1]
    \STATE Generate a volume of data samples (trajectories, paths, etc.) according to some specified mechanism, using an existing distribution of \acp{RV}.\label{step:basic_generate}
    \STATE Update \ac{RV} distributions using an elite subset of scored samples generated in Step~\ref{step:basic_generate}, to produce improved distributions for the next iteration.
\end{algorithmic}
\end{algorithm}
\end{figure}
In order to implement Algorithm~\ref{alg:basic_ce}, three essential factors must be defined:
\begin{enumerate}
    \item The family of probability distributions functions $\{f(\cdot;\mathbf{v})\}$ which is used to generate the random samples
    \item The method of scoring the samples in order to determine which are ``better''
    \item The updating rules for the parameters (based on cross-entropy minimization)
\end{enumerate}
With the above three ingredients supplied, the general algorithm can be applied to any discrete and continuous optimization problem by implementing Algorithm~\ref{alg:ce_opt}.  This method is referred to as  importance sampling because the probability distribution parameter update (Algorithm~\ref{alg:ce_opt}, Step~\ref{step:opt_update}) is based on an elite subset of samples which are deemed to be the most ``important'' ((Algorithm~\ref{alg:ce_opt}, Step~\ref{step:elite}).
\begin{figure}[!ht]
\begin{algorithm}[H]
\caption{Cross-Entropy Algorithm for Optimization}
\label{alg:ce_opt}
\begin{algorithmic}[1]
    \STATE Initialize parameter vectors ($\mathbf{u}$) for each random initial distribution
    \STATE Define $\widehat{\mathbf{v}}_0 = \mathbf{u}$.  Set $t=1$.
    \STATE Generate N samples $\mathbf{X_1,\ldots,X_N}$ from the distributions $f(\cdot;\widehat{\mathbf{v}}_{t-1})$ \label{step:opt_generate}
    \STATE Calculate the performance $S(\mathbf{X}_i)$ for all i, order them from smallest to largest: $S_{(1)} \leqslant \ldots \leqslant S_{(N)}$
    \STATE Compute the sample $(1-\rho)\!-\!quantile$ as: $\widehat{\gamma_t} = S_{(\lceil (1-\rho) N \rceil)}$ (i.e., the \textit{gamma element}) as well as the \textit{elite set} which consists of the set of vectors satisfying $S(\mathbf{X}_i) \geqslant \widehat{\gamma_t}$\label{step:elite} 
    \STATE Update the parameter vector $\widehat{\mathbf{v}}_{t-1}$ to $\widehat{\mathbf{v}}_t$ using the \textit{elite set}\label{step:opt_update} and optional smoothing parameter $\alpha$: \[\widehat{\mathbf{v}}_t = \alpha \widehat{\mathbf{v}}_t + (1-\alpha) \widehat{\mathbf{v}}_{t-1}\]
    \STATE If for some $t \geq d$, say $d = 3$, \[\widehat{\gamma_t} = \widehat{\gamma}_{t-1}= \ldots = \widehat{\gamma}_{t-d}\] then stop, otherwise, set $t = t+1$ and reiterate from Step~\ref{step:opt_generate}
\end{algorithmic}
    \textbf{Note:} The initial vector $\mathbf{u}$, the sample size $N$, the stopping parameter $d$, and the number $\rho$ have to be specified in advance, but the rest of the algorithm is ``self-tuning''.
\end{algorithm}
\end{figure}

The foundation of the \ac{CE} method  is to estimate the optimal importance sampling \ac{PDF} $g^*$ by choosing from a parametric class of \acp{PDF} $f(\cdot;\mathbf{v})$. The CE estimation minimizes the distance between the  the optimal \ac{PDF} $g^*$ and that of $f$. \cite{botev_2013}.  
The \ac{KL} divergence between $g^*$ and $f$ is given by:
\begin{align}\label{eq:KL_divergence}
    \mathcal{D}_{KL}(g^*,f) & =\int g^{*}(\mathbf{x}) \ln\frac{g^{*}(\mathbf{x})}{f(\mathbf{x})} d\mathbf{x} = \mathbb{E}\left[\ln\frac{g^{*}(\mathbf{X})}{f(\mathbf{X})}\right], \mathbf{X}\sim g^* \nonumber \\
    & = \int g^{*}(\mathbf{x}) \ln g^{*}(\mathbf{x})d\mathbf{x} - \int g^{*}(\mathbf{x}) \ln f(\mathbf{x})d\mathbf{x}
\end{align}

In order to minimize $\mathcal{D}_{KL}$, we need to maximize the right hand side of Equation~\ref{eq:KL_divergence}:

\begin{equation}
    \max_\mathbf{v} \int g^{*}(\mathbf{x}) \ln f(\mathbf{x;v})d\mathbf{x}
\end{equation}

Using importance sampling space, we get:

\begin{align}
    &\max_\mathbf{v} \int \mathrm{I}_{\{S(\mathbf{X}) \geqslant \gamma\}}\ln f(\mathbf{x;v})d\mathbf{x} \nonumber \\
    = &\max_\mathbf{v} \mathbb{E}_\mathbf{u} \mathrm{I}_{\{S(\mathbf{X}) \geqslant \gamma\}}\ln f(\mathbf{X;v})
\end{align}

Using change of measure (i.e., from $\mathbf{u}$ to some $\mathbf{w}$), the \ac{CE} method is reduced to a search for the optimal reference parameter vector, $\mathbf{v}^*$ of $g^*$, through \ac{CE} minimization:

\begin{align}\label{eq:ce_min}
\mathbf{v}^* & = \argmin_\mathbf{v} \mathcal{D}_{KL}(g^*,f(\cdot;\mathbf{v})) \nonumber \\
& = \argmax_\mathbf{v} \mathbb{E}_{\mathbf{u}} \mathrm{I}_{\{S(\mathbf{X}) \geqslant \gamma\}} \ln f(\mathbf{X};\mathbf{v}) \nonumber \\ 
& = \argmax_\mathbf{v} \mathbb{E}_{\mathbf{w}} \mathrm{I}_{\{S(\mathbf{X}) \geqslant \gamma\}} \ln f(\mathbf{X};\mathbf{v}) \frac{f(\mathbf{X;u})}{f(\mathbf{X;w})} 
\end{align}

where the subscript in the expectation operator indicates the density of $\mathbf{X}$.
This $\mathbf{v}^*$ can be estimated via the stochastic counterpart of Equation~\ref{eq:ce_min}:

\begin{equation}\label{eq:parameter_estimation}
    \widehat{\mathbf{v}} = \argmax_\mathbf{v} \frac{1}{N} \sum^N_{k=1} \mathrm{I}_{\{S(\mathbf{X}_i) \geqslant \gamma\}} W(\mathbf{X}_k;\mathbf{u,w}) \ln f(\mathbf{X}_k;\mathbf{v}) 
\end{equation}
\quad{where $W(\mathbf{X}_k;\mathbf{u,w}) = \frac{f(\mathbf{X}_k;\mathbf{u})}{f(\mathbf{X}_k;\mathbf{w})}$}

The optimal parameter $\widehat{\mathbf{v}}$ in Equation~\ref{eq:parameter_estimation} can often be obtained in explicit form, in particular when the class of sampling distributions forms an exponential family (e.g., exponential, normal, categorical, and multinomial)\cite{botev_2013}, as discussed in the next two sections.

\subsubsection{Optimal Parameter Estimation for Normal Distributions}
If \acp{RV} $\mathbf{X}_1,\ldots,\mathbf{X}_N$ are drawn from $\mathbf{X} \sim \mathcal{N}(\widehat{\mu}_{t-1},\widehat{\sigma}^2_{t-1})$, then $\mathbf{X}_1,\ldots,\mathbf{X}_N$ can be used to update the parameters $\widehat{\mu}_{t}$ and $\widehat{\sigma}^2_{t}$ using the following formulas \cite{botev_2013}:

\noindent{Let $N^e = \lceil \rho \cdot N \rceil$ (i.e., the number of items in the elite set) and let $\mathcal{I}$ be the indices of $N^e$.}\\
\noindent{For all $j=1,\ldots,n$ let}
\begin{equation}\label{eq:parameter_updating_norm_mean}
    \widehat{\mu}_{t,j} = \frac{1}{N^e} \sum_{k\in \mathcal{I}} W(X_{k,j};u, w_{t-1}) \cdot X_{k,j}
\end{equation}
\begin{equation}\label{eq:parameter_updating_norm_variance}
    \widehat{\sigma}^2_{t,j} = \frac{1}{N^e} \sum_{k\in \mathcal{I}} (W(X_{k,j};u, w_{t-1}) \cdot X_{k,j} - \widehat{\mu}_{t,j})^2
\end{equation}
\qquad\qquad{where $w_{t-1} = \langle {\mu}_{t-1},{\sigma}^2_{t-1} \rangle$}

\subsubsection{Optimal Parameter Estimation for Multinomial Distributions}
The multinomial distribution models the outcome of $n$ experiments where the outcome of each experiment is a categorical distribution.  A widely used example is where $(X_1,\ldots,X_n)$ has independent components such that $X_i=j$ with probability $v_{i,j}, i=1,\ldots,n, j=1,\ldots,m$.

If \acp{RV} $\mathbf{X}_1,\ldots,\mathbf{X}_N$ are drawn from $\mathbf{X} \sim Mult(\widehat{v}_{t-1,i,j})$, then $\mathbf{X}_1,\ldots,\mathbf{X}_N$ can be used to update the parameter vector $\widehat{v}_{t,i,j}$ using the following formula:

\begin{equation}\label{eq:parameter_updating_mult}
    \widehat{v}_{t,i,j} = \frac{1}{N^e} \sum_{k\in \mathcal{I}} {\mathrm{I}_{\{X_{k,i}=j\}}}
\end{equation}

Equation~\ref{eq:parameter_updating_mult} simply describes the fact that the updated probability $\widehat{v}_{t,i,j}$ is the number of elite samples for which the \textit{i}-th component is equal to \textit{j}, normalized by the number of samples \cite{botev_2013}.

\subsection{Machine Learning for Multivariate Time Series Data}\label{subsec:ml_time_series}
Most \acp{CPS} produce telemetry logs over the course of their operation.  From those logs we extract execution traces which can be represented as \ac{MTS}, time series data consisting of both discrete and continuous variables.  There are multiple models which are well-suited to classifying \ac{MTS}.  

\Acp{RDF} \cite{maharaj_2019} have become one of the most popular \ac{ML} algorithms due to the fact that they perform well over a wide range of applications, are easy to tune, and are fast to compute.  They have been widely applied to tabular data, but are also well-suited to time series data \cite{lazzeri_2020}.  Other authors have suggested modifications to the traditional \ac{RDF} algorithm for constructing time-series forest classifiers which split the data into random intervals, extract summary features, train a decision tree on the extracted features, and repeat those steps until the required number of trees are built.  As with the traditional structure, time series are classified according to majority vote \cite{maharaj_2019}.

In addition to \acp{RDF}, \acp{SVM} \cite{maharaj_2019} have also been applied to time series data classification and regression problems.  An \ac{SVM} is a classifier which constructs a separating hyperplane to distinguish between known groupings by maximizing the margin of the training data through the use of a kernel function.  In tabular data, that kernel function is often based on dissimilarity measures such as Euclidean distance (e.g., the radial basis function kernel).  Such kernels are typically not well suited to time-series data, and so \ac{DTW} is more often used to obtain the distance between two time series due to the alignment problems inherent in Euclidean distance measurements.  The Gaussian \ac{DTW} kernel, Gaussian elastic metric kernel, and global alignment kernel are examples of kernels well-suited to operate on time series \cite{maharaj_2019}.

\subsubsection{Deep Learning Approaches}\label{subsec:dl_approaches}
\Acp{RNN} \cite{lazzeri_2020} were first proposed in the 1980s, but have become more popular in recent years due to increased computational power.  They are a form of \ac{DNN} where neurons maintain information about previous inputs, allowing them to extract temporal patterns from data.  \Ac{RNN} models can suffer from \textit{vanishing gradients}, a phenomenon that results from the gradient value sometimes becoming so small that it does not contribute to the learning process, resulting in the model excluding long-term correlations from the beginning of the temporal sequence.  \Ac{LSTM} networks, a type of \ac{RNN}, are capable of learning long-term correlations through the addition of \textit{gates} and \textit{cells} which help prevent important information from being lost during the computation of gradient values in the backpropagation process.  A well-known drawback of \acp{DNN} and \acp{RNN} is their long training time compared to that of \acp{SVM} and especially that of \acp{RDF}.

\section{Method: Data Generation for Machine-Learned Correctness Property Training and Testing using \textit{HybridPair} Cross-Entropy Search}\label{sec:method}
To demonstrate the efficacy of this approach, we present an experimental method that: (1) automatically searches for correct and incorrect behavior, (2) automatically generates a large dataset of execution traces from those behaviors and (3) trains a \ac{RDF} to learn a model for the behavior of interest.  We call this approach \textit{HybridPair} \ac{CE} search: \textit{Hybrid} because it optimizes a probability distribution which is a hybrid of multiple underlying distributions, and \textit{Pair} because it searches for a pair of hybrid-paths, one of which exhibits correct behavior, the other of which does not.  We describe our approach to generating data using the \ac{CE} method (Section~\ref{subsec:data_gen}) and for training and testing the resulting \acp{MLCP} (Section~\ref{subsec:machine_learning}).

\subsection{Data Generation}\label{subsec:data_gen}
We simulate the behavior of an \textit{ego} vehicle (the \ac{AV}) in the presence of one or more \textit{adversary} vehicles with the goal of identifying both correct behavior (i.e., behavior that satisfies some requirement) and incorrect behavior (i.e., behavior that violates some accepted understanding of correct behavior) for the training of \acp{MLCP}.  Generating such events through random sampling techniques such as Monte Carlo methods is highly ineffective because assuming a relatively mature \ac{AV}, incorrect behavior such as collisions are rare events.  Hence, we search for these events using a novel adaptation of the \ac{CE} method.

Overall, there are six novel parts to our method of \ac{CE}-based data generation for \acp{MLCP}.
\begin{enumerate}
    \item Search for explainable paths
    \item Excluding ``Suicide Manuever'' paths using \textsc{vanilla}/\textsc{perturbed} hybrid-path pairs
    \item \Ac{CE} of hybrid probability distributions
    \item Integration of data-set variance into the \ac{CE} cost function
    \item Use of a weighted cost function
    \item Generation of \textsc{rudimentary} and \textsc{variant} hybrid-paths
\end{enumerate}

Algorithm~\ref{alg:ce_data_gen} describes our adaptations to the \ac{CE} algorithm and Sections~\ref{subsubsec:explainable} through \ref{subsubsec:rudimentary_variant} explain the abovementioned items. In this algorithm, samples ($\mathbf{X}_i$) are scored according to a custom cost function $S(\mathbf{X}_i)$, which is the one part of the algorithm that is customized to solve the underlying optimization problem.  The cost function is customized to capture the requirement that will ultimately be represented by the \ac{MLCP}, guiding the search for scenarios which are correct/incorrect with respect to that requirement.

\begin{figure}[!ht]
\begin{algorithm}[H]
\caption{\textit{HybridPair} Cross-Entropy Algorithm for Data Generation}
\label{alg:ce_data_gen}
\begin{algorithmic}[1]
    \STATE Initialize hybrid distribution parameter vectors ($\mathbf{u}$) for each adversary \label{step:ce_data_initialize}
    \STATE Define $\hat{\mathbf{v}}_0 = \mathbf{u}$.  Set $t=1$.
    \STATE Generate a hybrid sample $\mathbf{X_1,\ldots,X_N}$ from the density $f_{hybrid}(\cdot;\mathbf{v}_{t-1})$\label{step:ce_data_generate} 
        , \textbf{SEE *[1]} 
    \STATE Calculate the performance $S(\mathbf{X}_i)$ for all i, order them from smallest to largest: $S_{(1)} \ldots \leqslant S_{(N)}$, \textbf{SEE *[2]} \label{step:ce_data_score}
    \STATE Compute the sample $\rho\!-\!quantile$ as: $\widehat{\gamma_t} = S_{(\lceil \rho \cdot N \rceil)}$ (i.e., the \textit{gamma element}) as well as the \textit{elite set} which consists of the set of vectors satisfying $S(\mathbf{X}_i) \leqslant \widehat{\gamma_t}$ 
    \STATE Update the parameter vector $\widehat{\mathbf{v}}_{t-1}$ to $\widehat{\mathbf{v}}_t$ using the \textit{elite set} and optional smoothing parameter $\alpha$, \textbf{SEE *[3]}: \[\widehat{\mathbf{v}}_t = \alpha \widehat{\mathbf{v}}_t + (1-\alpha) \widehat{\mathbf{v}}_{t-1}\] 
    \STATE If all gamma elements meet satisfying conditions then stop, otherwise, set $t = t+1$ and reiterate from Step~\ref{step:ce_data_generate} \label{step:ce_data_is_gamma_neg} 
\end{algorithmic}

    \textbf{*[1]}:   Each $\mathbf{X_i}$ is a sequence of positions and accelerations drawn from the hybrid distribution\\
    \textbf{*[2]}:   We calculate the elite set as $\mathbf{X_{S(1)},\ldots,X_{S(\rho \cdot N)}}$ instead of $\mathbf{X_{S((1-\rho)N)},\ldots,X_{S(N)}}$ (From Algorithm~\ref{alg:ce_opt}) because, in line with the approach taken by Falsification methods \cite{annpureddy_2011}, lower scores are considered better than higher scores. As such, $S(\mathbf{X}_i) < 0$ indicates that the sample (e.g. path) satisfies rigid constraints such as ``vehicle must not collide with ego''\\
    \textbf{*[3]}: We set $\alpha = 0.8$ in our implementation\\
\end{algorithm}
\end{figure}

\subsubsection{Search for Explainable Paths}\label{subsubsec:explainable}
The resulting model learned by any \ac{ML} algorithm is a direct result of the data used to train it.  Because our goal is to bootstrap the creation of \acp{MLCP} through data generation, we ensure the resulting model is explainable by focusing on the explainability of the training data we generate.  As such, we use real-world data collected by the \ac{CADMV}.  In order to enable the safe testing and deployment of \acp{AV} in California, in 2014 the \ac{CADMV} began requiring companies testing \acp{AV} on public roads to submit disengagement reports.  These reports record instances where the vehicle's autonomous control system is disengaged, either by the vehicle or by the human backup driver, and the \ac{CADMV} publishes on an annual basis the disengagement reports (as well as crash reports) from all companies testing \acp{AV}.  Disengagement reports are of particular interest because:
\begin{itemize}
    \item \Ac{AV} technology being tested on public roads is relatively mature, and hence unlikely to make lots of mistakes under normal driving conditions, so reported disengagements usually indicate the occurrence of critical/unprecedented scenarios.
    \item Disengagements represent a failure of human-machine teaming, a critical component in developing trust in the use of \acp{CPS} in areas such as automotive vehicles.  If the disengagement is initiated by the \ac{AV}, it indicates an error in the underlying \ac{CPS} which reduces human trust in the system.  If the disengagement is driver-initiated, it usually represents a preemptive takeover to prevent a potential perceived incorrect action by the \ac{CPS}, also indicating lack of trust.   
\end{itemize}

Therefore, when we look at what type of interactions for which \acp{MLCP} would provide the most benefit to establishing trust in the dependability of \acp{AV}, we use natural language descriptions from \ac{CADMV} reports as a starting point of a CE search for paths of a \ac{MLCP} dataset.  For example, on September 24, 2019, the \ac{AV} company Zoox reported a disengagement by a backup driver due to ``Prediction discrepancy; incorrect yield estimation for a vehicle cutting into the lane of traffic'' \cite{CADMV_2022}.  Since our \ac{MLCP} dataset generation method uses such natural language descriptions as its starting point, the resulting dataset is explainable.  In the example used by our experimentation, data generated for the behavior of an \ac{AV} interacting with nearby vehicles performing lane change maneuvers will induce an explainable dataset for one or more resulting explainable \acp{MLCP}.  This is the example we take for our experimentation.

The essence of this paper is the application of novel extensions to the \ac{CE} search algorithm (Algorithm~\ref{alg:ce_data_gen}).  One part of that extension resides in the \ac{CE} cost function (Algorithm~\ref{alg:ce_data_gen}, Step \ref{step:ce_data_score}) which is written to detect paths that conform to, and paths that do not conform to, the particular \ac{MLCP} disengagement instance.

In addition, we assume that the ego and a single adversary vehicle are not the only vehicles on the road.  We add an additional vehicle to the simulation which we call the \textit{independent adversary} (\textit{independent} because its behavior is controlled by a probability distribution that is independent of the adversary's) to both increase explainability and demonstrate the generalizability of our approach to complex multi-agent systems with potentially hundreds of adversaries or more.

\subsubsection{Excluding ``Suicide Manuever'' Paths using \textsc{vanilla}/\textsc{perturbed} Hybrid-Path Pairs}\label{subsubsec:vanilla_pert}
In the simulation-based search for correct and incorrect behavior, we attempt to rule out the types of trivially unsafe behavior that is not likely to be seen in real traffic scenarios or is not useful for generating \acp{CP}.  For example, simply making a vehicle perform a high-speed ``suicide'' collision with an \ac{AV} travelling predictably on a highway is clearly a collision scenario, but is not an event of interest for \acp{MLCP} because there is little the \ac{AV} can do about it.  Therefore, we take the philosophical approach that collisions (or near-misses) of interest are the result of only slight variations from otherwise acceptable behavior.  The ``normal'' behavior that does not result in a collision (or violation of another condition of interest) we call a \textsc{vanilla} hybrid-path of the adversary.  The behavior that is similar to the correct path, but does result in a violation, we call the \textsc{perturbed} hybrid-path of the adversary.  A pair of such \textsc{core} paths of the adversary (\textsc{vanilla}/\textsc{perturbed}) can be visualized on a two-dimensional decision boundary as depicted in Fig.~\ref{fig:decision_boundary}.  In reality, the separating hyperplane is of a significantly higher dimension. This boundary is an abstract border separating the path parameters of a collision-avoiding vehicle and those of a slightly perturbed collision-inducing path for the same vehicle.

Both \textsc{vanilla} and \textsc{perturbed} paths of the adversary are hybrid-paths, i.e., they are sequences of location (a Categorically distributed value) and acceleration (a Normally distributed value).  The concept of a \textsc{vanilla}/\textsc{perturbed} path pair (i.e., showing incorrect behavior that is subtly different from correct behavior) helps to support both the realism and explainability of the data ultimately used to train the \acp{MLCP}.  The \textsc{vanilla}/\textsc{perturbed} path pairs make up the \textit{Pair} component of the \textit{HybridPair} term we use to refer to our \ac{CE} search technique.

We adapted the \ac{CE} search algorithm to search for \textsc{vanilla}/\textsc{perturbed} pairs of adversary paths by adding the following constraints to the \ac{CE} cost function:
\begin{itemize}
    \item \textsc{vanilla} adversary path has no collision with ego
    \item \textsc{perturbed} adversary path has collision with ego
    \item Distance between \textsc{vanilla} and \textsc{perturbed} adversary paths (as measured by a combination distance metric of location and acceleration) must be below some threshold
\end{itemize}

\subsubsection{Cross Entropy Search Using A Hybrid Probability Distribution}\label{subsubsec:hybrid}
Typically, the \ac{CE} method focuses on drawing samples from a single underlying \ac{PDF} for the simulation of rare events or solving a combinatorial optimization problem \cite{rubinstein_2004}.  In our case, we take the novel approach of optimizing a \textit{hybrid} probability distribution consisting of the following underlying distributions:
\begin{itemize}
    \item Multinomial (Categorical) Distribution: This distribution is used to represent the adversary's position
    \item Gaussian (Normal) Distribution: This distribution is used to represent the adversary's speed/acceleration
\end{itemize}
By using a hybrid distribution, \ac{CE} can search through a wide range of paths that are variable in both position and speed, allowing the resulting paths to encompass the full range of driving behavior in order to find rare paths that may lead to collisions or near misses.  The suggested approach is not the same as searching for solutions that optimize one distribution and then searching, with that set, for solutions that optimize the other distribution. Rather, \ac{CE} samples are drawn from a combination of both distributions, and the \ac{CE} cost function is a hybrid cost function, inducing a score used for creating a new set of hybrid distributions.

\subsubsection{Integrating Data-Set Variance Into the Cross Entropy Cost Function}\label{subsubsec:variance}
In Algorithm~\ref{alg:ce_data_gen}, each batch of \ac{CE} (i.e., a certain number of \ac{CE} iterations that occur prior to the $\hat{\gamma}_t < 0$) results in a single pair of hybrid-paths of the adversary - a \textsc{perturbed} (which will serve as a 1-labeled path for the \ac{MLCP}) and a \textsc{vanilla} path (which will serve as a 0-labeled path for the \ac{MLCP}).  Because we have a goal of ensuring the \ac{MLCP} is trained on a wide variety of data to maximize generalizability, we integrate the variance between the existing paths in the dataset and the newly generated \textsc{vanilla}/\textsc{perturbed} pair of adversary paths into the cost function, $S(\textbf{X}_i)$.  Given that collisions can happen in a variety of different ways, the modified cost function incentivizes \ac{CE} to discover collisions that differ from previous collisions to ensure that a variety of paths are represented in the training data.

Specifically, this process is comprised of three steps:
\begin{enumerate}
    \item Determine the current distribution of \textsc{vanilla} adversary paths in the dataset, and the current distribution of  \textsc{perturbed} adversary paths in the dataset.  Each of those distributions is the combination of two normal distributions.
    
    Distribution of \textsc{Vanilla} paths of the adversary:
    \begin{itemize}
        \item $\mathcal{N}_{\textsc{vanilla},location}(\mu_{\textsc{vanilla},location},\sigma^2_{\textsc{vanilla},location})$
        \item $\mathcal{N}_{\textsc{vanilla},accel}(\mu_{\textsc{vanilla},accel},\sigma^2_{\textsc{vanilla},accel})$
    \end{itemize}

    Distribution of \textsc{Perturbed} paths:
    \begin{itemize}
        \item $\mathcal{N}_{\textsc{perturbed},location}(\mu_{\textsc{perturbed},location},\sigma^2_{\textsc{perturbed},location})$
        \item $\mathcal{N}_{\textsc{perturbed},accel}(\mu_{\textsc{perturbed},accel},\sigma^2_{\textsc{perturbed},accel})$
    \end{itemize}
    
    \item Calculate  metrics for the candidate \textsc{vanilla} and \textsc{perturbed} hybrid-paths of the adversary.  The metrics map the candidate hybrid-path's location vector ($\mathbf{X}_{loc} \sim Mult(\widehat{v}_{t-1,i,j})$)
   to a scalar value ($\chi_{location}$) representing the hybrid-path's entire location vector; likewise, 
   map the candidate hybrid-path's acceleration vector ($\mathbf{X}_{accel} \sim \mathcal{N}(\widehat{\mu}_{t-1},\widehat{\sigma}^2_{t-1})$) to a scalar value $\chi_{acceleration}$.
    \item Score paths as ``better'' (lower scores) for which ($\chi_{\textsc{VP},la}$) is closer to  $\mu_{\textsc{VP},la} \pm 2\cdot \sigma_{\textsc{VP},la}$, for all pairs $\textsc{VP} \in \textsc{VANILLA},\textsc{PERTURBED}$, $la \in location, acceleration$.  In other words, ensure the newly discovered hybrid-paths are close to two standard deviations away from the current mean of all previously discovered hybrid-paths.  
\end{enumerate}
    Scoring paths whose combined location and acceleration metrics are closer to $\mu \pm 2\cdot\sigma$ from all previous paths incentivizes the \textit{HybridPair} \ac{CE} search to add paths to the dataset that vary from previous paths, increasing the variance of the dataset.

\subsubsection{Weighted Cost Function}\label{subsubsec:weighted}
The \textit{HybridPair} \ac{CE} algorithm searches for a \textsc{vanilla}/\textsc{perturbed} pair of adversary paths using a cost function, $S(\textbf{X}_i)$, that accounts for five factors:
\begin{enumerate}
    \item Categorical cost ($\mathcal{C}$): prioritize shorter paths
    \item Gaussian cost ($\mathcal{G}$): ensure the normally distributed speed values are positive
    \item Distance between \textsc{vanilla} and \textsc{perturbed} hybrid-paths of the adversary ($\mathcal{D}_{v/p}$): ensure that these paths are similar according to the philosophy discussed in Section \ref{subsubsec:vanilla_pert}
    \item Variance ($\mathcal{V}$): ensure that the next hybrid-path added to the dataset maximizes the distance from previously discoverd hybrid-paths (as discussed in Section \ref{subsubsec:variance})
    \item Rigid Constraints ($\mathcal{R}$): a hybrid-path must obey certain rigid constraints (e.g., requiring a \textsc{perturbed} (\textsc{vanilla}) path of the adversary to collide (not collide) with the \textit{ego} vehicle, requiring all paths to avoid obstacles).  Failure to conform to a rigid constraint results in a poor score.
\end{enumerate}

The first four factors above can each be assigned different weights according to the prioritization placed on each component of the score (Equation~\ref{eq:weighted_cost}).

\begin{equation}\label{eq:weighted_cost}
    S(\mathbf{X}_i) = \alpha_1 \cdot \mathcal{C} + \alpha_2 \cdot \mathcal{G} + \alpha_3 \cdot \mathcal{D}_{v/p} + \alpha_4 \cdot \mathcal{V} + \mathcal{R}
\end{equation}

For example, placing a higher weight on the categorical cost ($\alpha_1$) indicates a preference for shorter paths over and above the other factors.  Similarly, a lower weight ($\alpha_3$) for the \textsc{vanilla}/\textsc{perturbed} distance (which is a smaller negative number the closer they are to each other) would indicate a lower emphasis on the similarity between the \textsc{vanilla} (non-collision) and \textsc{perturbed} (collision) hybrid-paths of the adversary.  The final factor (rigid constraints) is not weighted  because it is either zero (all rigid constraints satisfied) or it is an extremely high value that dwarfs all the other factors, indicating that rigid constraints were violated.

Through experimentation, we utilize the following weights: \[\alpha_1 = 3, \alpha_2 = 2, \alpha_3 = 2, \alpha_4 = 1\]

\subsubsection{\textsc{Rudimentary} and \textsc{Variant} Hybrid-Paths}\label{subsubsec:rudimentary_variant}
\begin{figure}[ht] 
\centering
\includegraphics[width=0.7\columnwidth]{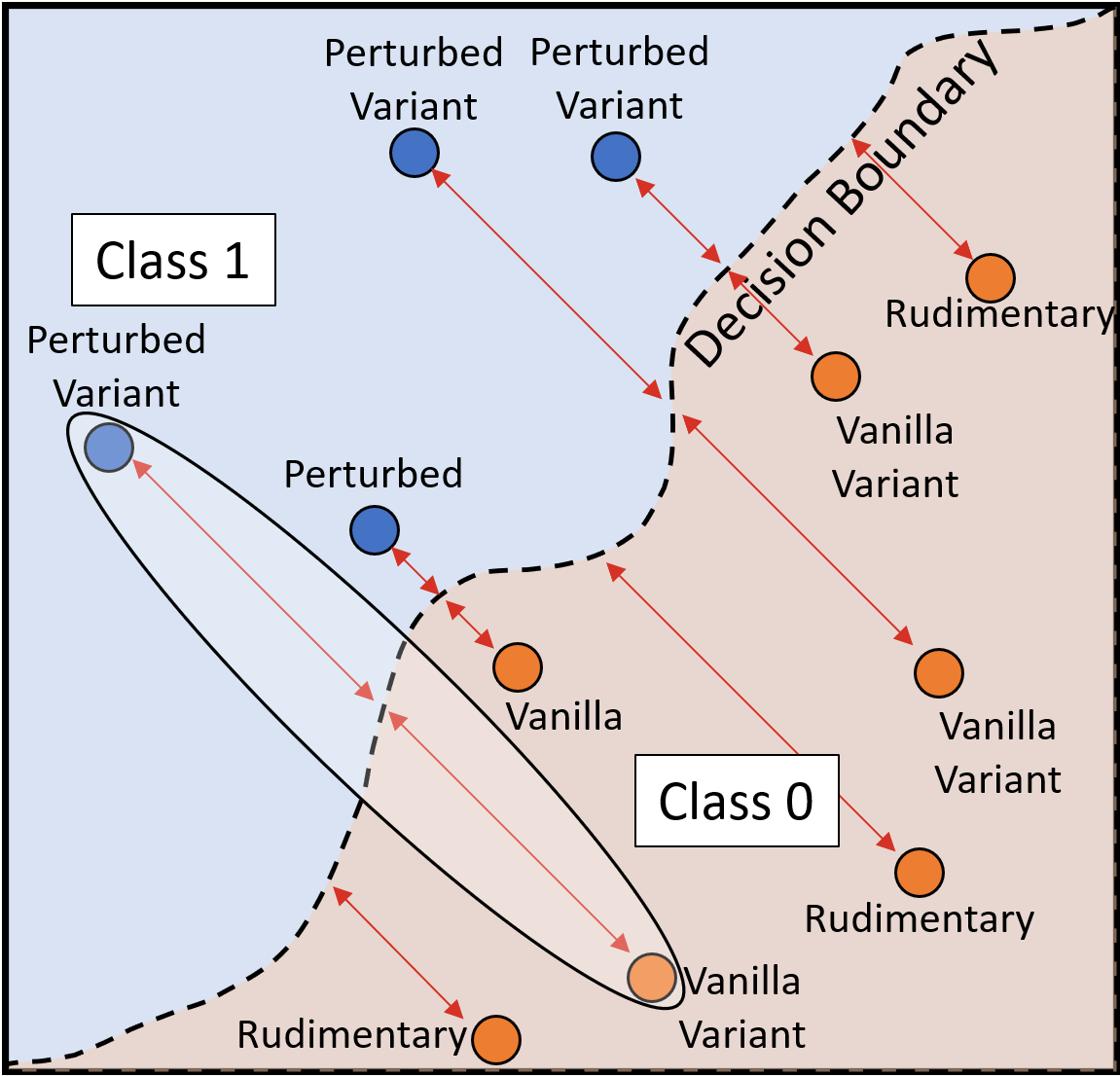}
\caption[]{A visual depiction of the execution traces used to train the model.  Note the proximity of the \textsc{vanilla}/\textsc{perturbed} hybrid-pairs to the decision boundary, and the larger relative distance of the \textsc{variant} and \textsc{rudimentary} traces.}
\label{fig:decision_boundary}
\end{figure}

The \textit{HybridPair} \ac{CE} search can be seen as finding the boundary between 1-labeled (\textsc{perturbed}) and 0-labeled (\textsc{vanilla}) \ac{ML} data items (See Fig.~\ref{fig:decision_boundary}).  Although each component of the pair has different labels, its constituent hybrid-paths are very similar to each other (see Fig.~\ref{fig:path_vis_trunc_labeled}), which is a direct result of the algorithm's purpose of generating explainable paths in the form of \textsc{vanilla}/\textsc{perturbed} pairs of adversary paths.

A classifier however, typically needs to be trained using additional data items, such as 0-labeled data items that are farther away from the decision boundary (e.g., a vehicle traveling along the same sequence of locations as the \textsc{vanilla} adversary path, yet whose sequence of accelerations are sufficiently different).  Hence, we generate two additional types of hybrid-paths: \textsc{rudimentary} paths and \textsc{variant} paths.

Fig.~\ref{fig:decision_boundary} depicts a pair of \textsc{vanilla}/\textsc{perturbed} \textit{core} paths, as well as three additional  types of adversary hybrid-paths called \textit{helper} paths, which are used for generating a robust \ac{MLCP} training set. More specifically, the path types are:
\begin{itemize}
\item \textit{Core} Adversary Hybrid-Paths
    \begin{itemize}
        \item \textsc{vanilla}: Class 0.  Correct behavior that would be incorrect with a slight perturbation.
        \item \textsc{perturbed}: Class 1. Incorrect behavior that would be correct without a slight perturbation.
    \end{itemize}
\item \textit{Helper} Adversary Hybrid-Paths
    \begin{itemize}
        \item \textsc{variant vanilla}: Class 0. Correct behavior farther from the decision boundary.
        \item \textsc{variant perturbed}: Class 1. Incorrect behavior farther from the decision boundary.
        \item \textsc{rudimentary}: Class 0.  Correct behavior that represents the normal, routine behavior of a system.
    \end{itemize}
\end{itemize}

\textsc{Rudimentary} paths are those generated from distributions that have not been iteratively updated by \ac{CE}.  They represent the types of naturalistic non-collision paths likely to be prevalent in randomly sampled real-world data and they are like paths that would be generated from pure Monte Carlo simulation.  \textsc{Variant} paths are generated by the same pair of distributions that \ac{CE} computed for \textsc{vanilla}/\textsc{perturbed} hybrid-path pairs (discussed in Section~\ref{subsubsec:vanilla_pert}), but they are sufficiently different than their \textsc{vanilla}/\textsc{perturbed} counterparts.  One such \textsc{variant} hybrid-path pair is shown in Fig.~\ref{fig:decision_boundary} to illustrate their larger relative distance from the decision boundary.

\subsection{Machine Learning}\label{subsec:machine_learning}
In order to test the ability to train \ac{ML} models from the generated data, we first pre-process the data by taking the following steps:
\begin{enumerate}
    \item For each \textsc{vanilla}/\textsc{perturbed} hybrid-pair of adversary paths, identify the time at which the collision between the adversary and the ego occurs in the \textsc{perturbed} path.  This time is denoted as $t_{collision}$.\label{van_per}
    \begin{enumerate}
        \item Starting at $t_{collision}$, go back in time some number of seconds ($X$) prior to the collision ($t_{collision-X}$).\label{1a}
        \item Then take a series of time-sequenced features at some sampling rate ($R$) for some number of seconds ($Y$) before $t_{collision-X}$ (i.e., $t_{collision-(X+Y)}$).\label{1b}
        \item Features taken from the \textsc{vanilla} hybrid-path of the adversary are taken at exactly the same intervals and frequency as from the \textsc{perturbed} hybrid-path of the adversary.
    \end{enumerate}
    \item For each \textsc{Rudimentary} and \textsc{Variant Vanilla} hybrid-path of the adversary, take features from the exact same interval and frequency as described above in Step~\ref{van_per}.
    \item For each \textsc{Variant Perturbed} hybrid-path, determine $t_{collision}$ and then then follow Steps~\ref{1a} and \ref{1b}.
\end{enumerate}
For example, if the collision occurred at $t=5 s$, you may choose to set $X=1$, $Y=2$, and $R=0.5$.  This would mean that the features taken from the \textsc{vanilla}/\textsc{perturbed} adversary paths would be a sequence of features starting at $t=2 s$ and ending at $t=4 s$, sampled every $0.5s$.   The \textsc{variant vanilla} and \textsc{rudimentary} hybrid-paths would have features sampled at the same indices.  The \textsc{variant perturbed} hybrid-paths would have features sampled relative to their $t_{collision}$.  If 33 features are collected at every timestep, then there would be 165 total features for each path; the row of 165 features taken from the \textsc{vanilla}, \textsc{variant vanilla}, and \textsc{rudimentary} paths are labeled as a '0', and the row of 165 features taken from the \textsc{perturbed} and \textsc{variant perturbed} hybrid-paths are labeled as a `1.'  The features used to train each \ac{ML} model include the relative position, velocity, and acceleration of the \textit{ego} and \textit{adversary} vehicles.

The \ac{ML} classifier is trained and tested using this dataset.  Consequently, it is trained to answer the question (with the above parameters), ``Given data from 3 seconds before the potential collision up to 1 second before the potential collision, can I predict whether or not there will actually be a collision between the adversary and the ego?''.

\section{Experiment}\label{sec:experiment}

\subsection{Implementation and Simulation}\label{subsec:implementation_simulation}
The technique and novel contributions described above are implemented in approximately 10k lines of code which consists of the following components:
\begin{enumerate}
    \item A \ac{CE} generator that implements the CE method and produces paths to be executed by the simulator in addition to tracking the state of the underlying distributions
    \item A scenario executor which takes the paths generated by \ac{CE} and executes/analyzes them in the simulator
    \item Data pre-processing algorithms
    \item \Ac{ML} model training and testing
\end{enumerate}

For our simulator we utilize \ac{CARLA} \cite{carla} due to its strong developer support and flexible API.  

\subsubsection{Simulation Example}\label{subsubsec:sim_example}

\begin{figure}[ht] 
\centering
\includegraphics[width=\columnwidth]{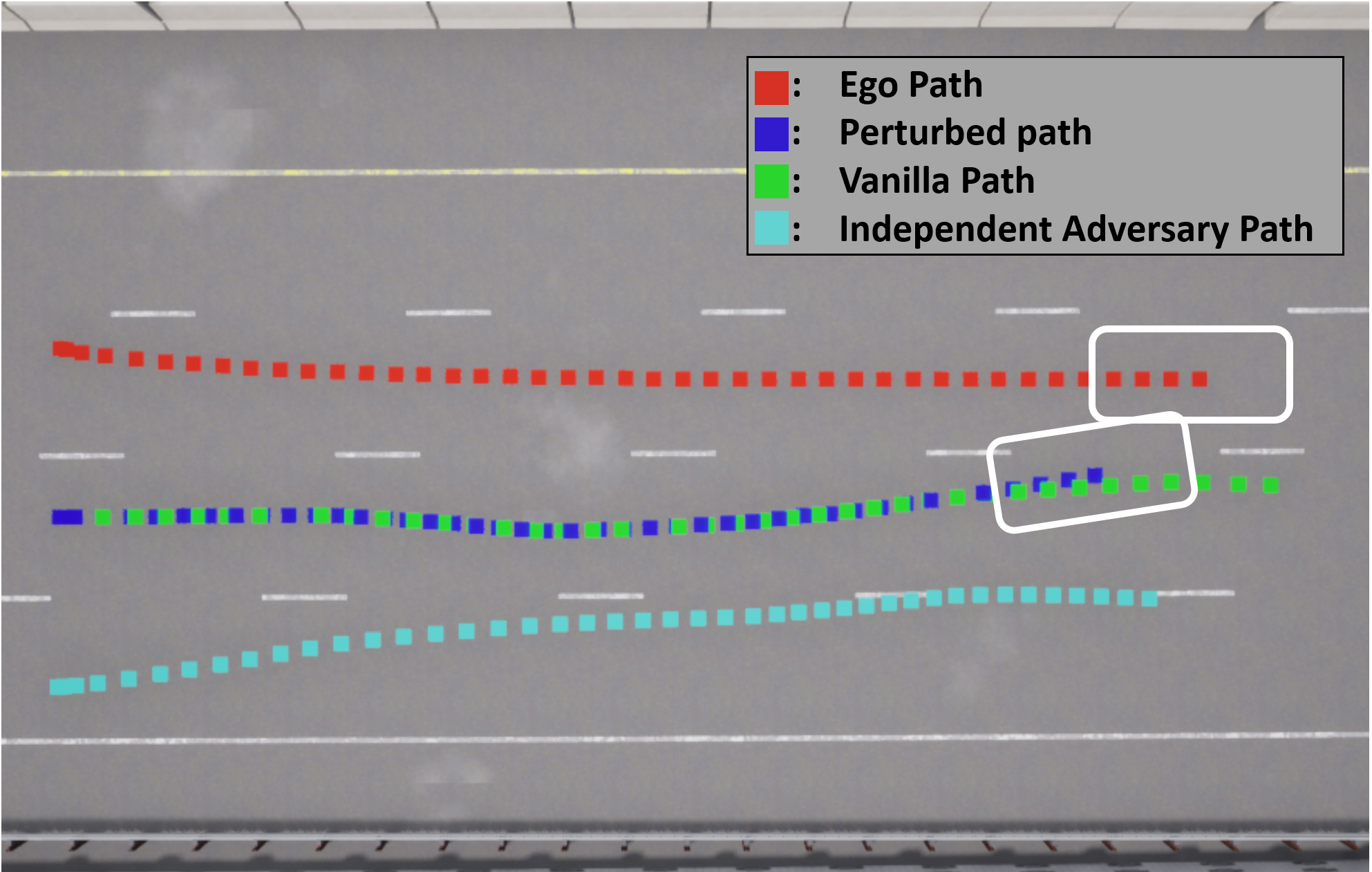}
\caption[Vanilla/Perturbation Path Pair-Position]{A visual depiction of a \textsc{vanilla}/\textsc{perturbed} paths with the ego and adversary trajectories.  Note the similarity in position of the \textsc{vanilla}/\textsc{perturbed} paths of the adversary prior to the collision.  The position of each vehicle is represented by grid coordinate, where each grid cell is approximately the length of a vehicle and the width of one half of a traffic lane.}
\label{fig:path_vis_trunc_labeled} 
\end{figure}

\begin{figure}[ht] 
\centering
\includegraphics[width=\columnwidth]{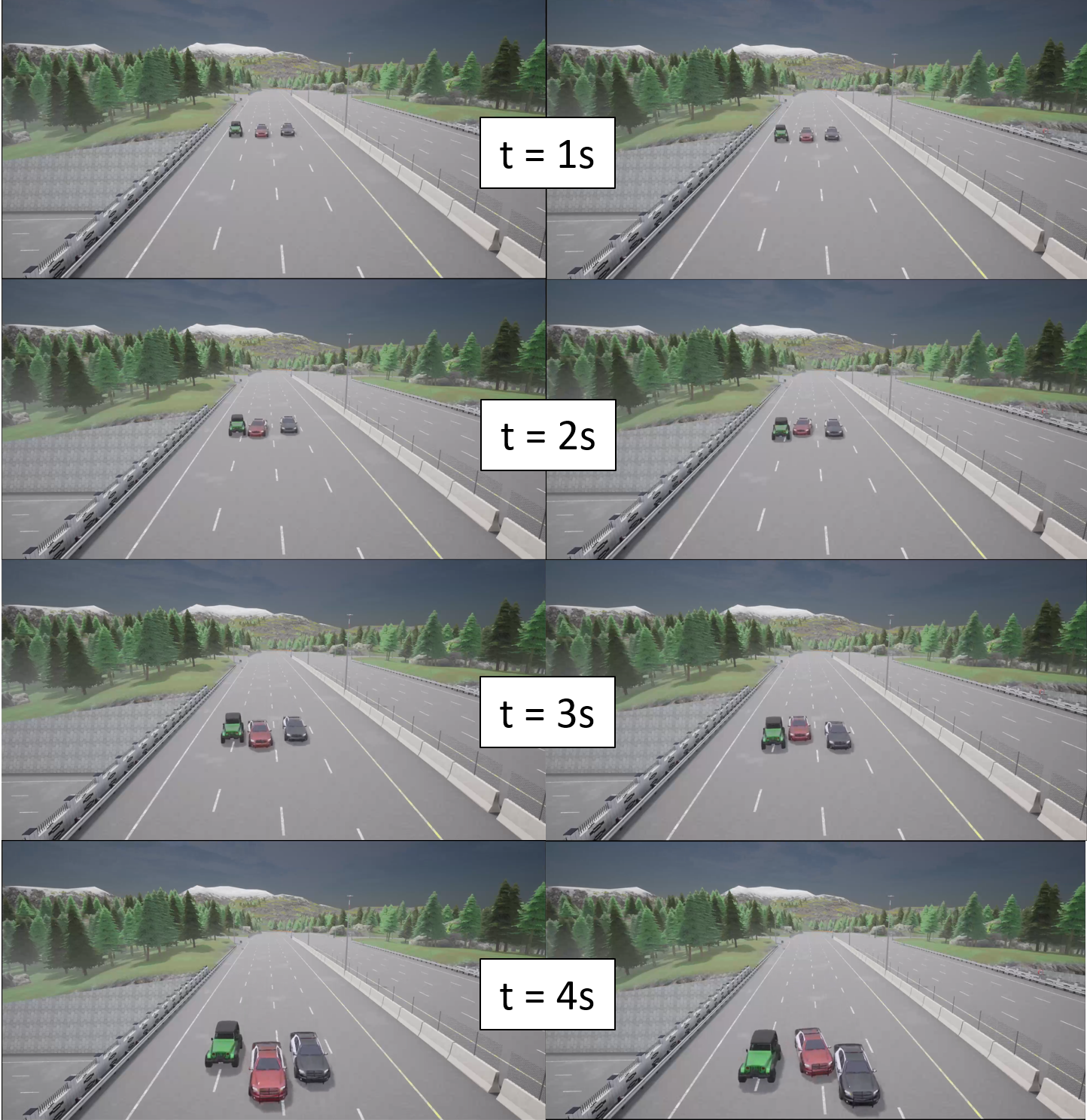}
\caption[]{The same scenario represented in Fig.~\ref{fig:path_vis_trunc_labeled} rendered in the simulator.  The left column shows the \textsc{vanilla} adversary path, and the right column shows a \textsc{perturbed} path, where the \textsc{perturbed} adversary path results in a collision with the ego vehicle at $t=4s$.}
\label{fig:sequence} 
\end{figure}

Each simulation involves three vehicles: the \textit{ego} vehicle (i.e, the \ac{AV}), the \textit{adversary} vehicle, and the \textit{independent adversary}.  The adversary's path either results in a collision with the ego, in which case it is a \textsc{perturbed} path, or it does not, making it a \textsc{vanilla} path.  The \textsc{independent adversary} causes the \textsc{perturbed} path of the adversary to result in a collision by encroaching into the \textit{adversary's} lane and forcing it into a collision with the \textit{ego}.  One such set of paths can be seen in Figs.~\ref{fig:path_vis_trunc_labeled} and \ref{fig:sequence}.  This visualization clearly depicts the similarity in position (the Multinomial part of the hybrid distribution discussed in Section \ref{subsubsec:hybrid}) between the \textsc{vanilla}/\textsc{perturbed} adversary paths prior to the collision. The \textsc{perturbed} path collides with the ego vehicle, hence its path is truncated at the point of collision (the bounding boxes of the ego and (\textsc{perturbed}) adversary vehicles at the point of collision are shown in Fig.~\ref{fig:path_vis_trunc_labeled}).   The \textsc{vanilla} path of the adversary does not collide with the ego vehicle and is truncated at the point that the adversary's \textsc{perturbed} path resulted in a collision.  Note how close the \textsc{vanilla}/\textsc{perturbed} paths are, a direct result of the \ac{CE} search approach previously discussed.

\subsection{Research Questions}\label{subsec:research_questions}
To validate this research, we aim to answer the following research questions.
\\\noindent{\textbf{Research Question 1:}} Can \ac{ML} be used to assert about critical properties of \acp{CPS} in their environment?
\\\noindent{\textbf{Research Question 1(a):}} Which \ac{ML} models perform best on these datasets?
\\\noindent{\textbf{Research Question 2:}} Can \acp{MLCP} serve as requirements for systems which must give advance notice to humans (e.g., alert a human of an imminent collision)?
\\\noindent{\textbf{Research Question 2(a):}} How does the performance of the \ac{MLCP} vary depending on desired detection time (i.e., the difference in time between when the \ac{MLCP} flags and when the potential collision occurs)? 
\\\noindent{\textbf{Research Question 3:}} What is the impact of adding \textsc{rudimentary} paths to the training set?
\\\noindent{\textbf{Research Question 4:}} What is the impact of adding \textsc{variant} paths to the training set?

\section{Results}\label{sec:results_analysis}

\subsection{Results for Research Question 1}\label{subsec:RQ1}
To answer this research question, we trained a classifier on the various paths discussed in Section~\ref{sec:method}, all of which are labeled as 0 (no collision) or 1 (collision).  The classifier learns a model which can distinguish between adversary lane changes that do result in collisions, and adversary lane changes that are collision-free.  A natural language equivalent for this model would be "Alert when a vehicle in the right adjacent lane is executing a lane change trajectory which will cause a collision with the \ac{AV}."  The \ac{MLCP} could be used to detect the condition discussed in Section~\ref{subsubsec:explainable} where \acp{AV} experienced disengagements due to ``Prediction discrepancy; incorrect yield estimation for a vehicle cutting into the lane of traffic'' \cite{CADMV_2022}.  We begin by training a \ac{RDF} on a dataset consisting of 770 \textsc{vanilla} adversary paths (including \textsc{vanilla} paths contained in \textsc{vanilla}/\textsc{perturbed} path pairs of the adversary, \textsc{rudimentary}, and \textsc{variant vanilla}) and 210 \textsc{perturbed }paths (including \textsc{perturbed} paths contained in \textsc{vanilla}/\textsc{perturbed} paths pairs and \textsc{variant perturbed} paths).

Because the distance between the ego and the nearest adversary is used as a feature for the \ac{ML} model, we do not sample features all the way up until the collision to avoid giving the classifier information indicating that a collision has already occurred (i.e. distance = 0).  Rather, we sample features up to one second prior to the collision.  Using the preprocessing steps described in Section~\ref{subsec:machine_learning}, we set $X=1$, $Y=2$, and $R=0.2$ which equates to taking two seconds of features every fifth of a second, starting three seconds prior and ending one second prior to the collision.

The dataset exhibits only a minor class imbalance; the imbalance is significantly smaller than a real-world dataset where 1's are even less frequent, which is a significant advantage of the \ac{CE} search.  Due to small imbalance, we chose not to perform any undersampling of the 0's or oversampling of the 1's; rather, we evaluate each classifier based primarily on its F1 score as well as its  balanced accuracy ($\frac{1}{2} \cdot (sensitivity + specificity)$ in order to use metrics most well-suited to imbalanced data.  Table~\ref{table:ML} demonstrates that such a condition can be detected with very high accuracy, showing that \acp{MLCP} can indeed be used to assert about critical properties of autonomous \acp{CPS}.

\begin{table}[H]
    \vspace{0mm} 
    \caption[]{Machine Learning Results}
    \centering
    \begin{tabular}{lccc@{\hspace{2em}}c} 
    \toprule 
    \textbf{Classifier} & \textbf{Recall} & \textbf{Precision} & \textbf{F1 Score} & \bfseries\makecell{Training\\Time (seconds)} \\ \midrule
    RDF & 0.99 & 0.95 & 0.97 & 0.25 \\
    SVM & 0.78 & 0.89 & 0.83 & 0.32 \\
    LSTM & 0.98 & 0.91 & 0.94 & 46.5 \\
    \bottomrule 
    \end{tabular}%
    \label{table:ML}
\end{table}

\subsection{Results for Research Question 1(a)}\label{subsec:RQ1a}
To determine the most appropriate model to use on the dataset, we compared the \ac{RDF} results against an \ac{SVM} (\ac{DTW} kernel) as well as an \ac{LSTM} for the same dataset.  Table~\ref{table:ML} shows the results from the various models, as well as their time to train.  We see that the \ac{RDF} is the most performant model, demonstrating high F1 score as well as the least time to train.  While the \ac{RDF} and \ac{SVM} were trained on a laptop (Intel$^{\circledR}$ Core i9-9980HK CPU @ 2.40GHz with 64GB of memory), the \ac{LSTM} required a High Performance Computing cluster with a NVIDIA V100 GPU due to memory requirements.  

\subsection{Results for Research Question 2}\label{subsec:RQ2}
In order to provide advance notice to a human (e.g., some sort of visual, auditory, or haptic feedback that human engagement is required), a \ac{MLCP} must first be able to detect that a condition will take place some amount of time before the condition actually occurs.  The exact amount of time required depends significantly on the driver, the vehicle, and the operating conditions at the time of takeover; those types of considerations are outside the scope of this work.  We demonstrate through the preprocessing steps previously described and the results listed in Table~\ref{table:ML} (i.e., one second of advance notice) that indeed, \acp{MLCP} can provide advance notice of future potentially hazardous conditions.  In Section~\ref{subsec:RQ2a} we investigate the accuracy of that prediction as the advance notice time varies.  The ability to predict incidents in advance, and quantify the uncertainty of those predictions, has been discussed in previously published work \cite{litton_2023}.

\subsection{Results for Research Question 2(a)}\label{subsec:RQ2a}
\begin{figure}[!ht] 
\centering
\includegraphics[width=\columnwidth]{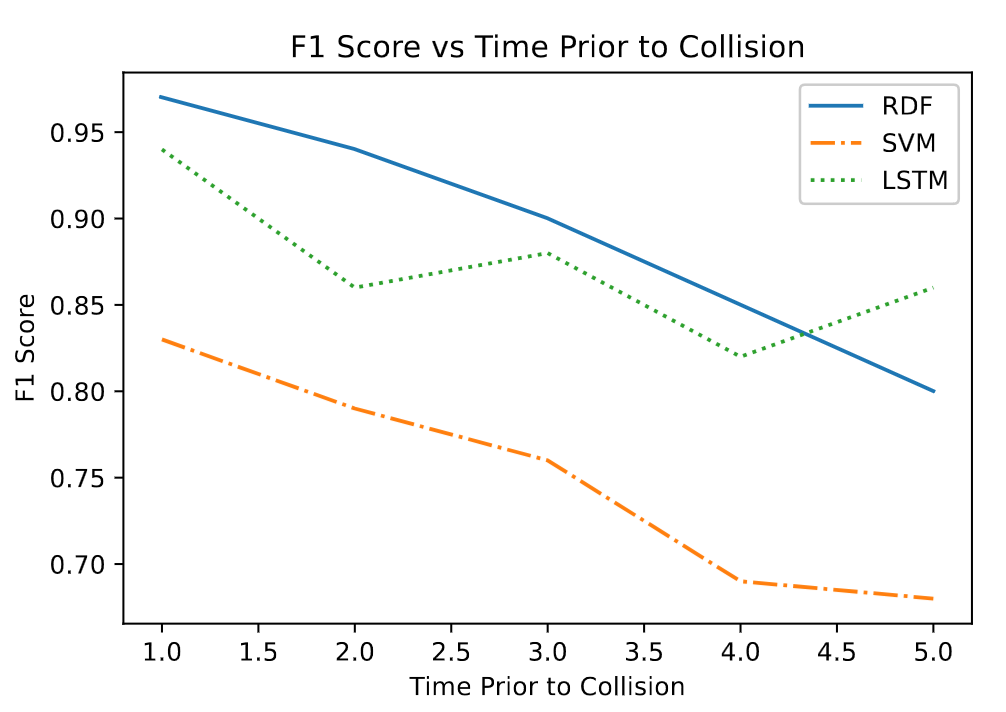}
\caption[TTC vs F1 Score]{The F1 score of the classifiers tested plotted against the time prior to the collision at which features were collected.}
\label{fig:ttc_vs_f1} 
\end{figure}

Fig.~\ref{fig:ttc_vs_f1} displays the performance (F1 score) of various classifier models as $X$ in $t_{collision-X}$ is varied from one second to five seconds (see Section~\ref{subsec:machine_learning} for an explanation of pre-processing steps and the calculation of $t_{collision-X}$).  As expected, the F1 score decreases (on average) with an increase in advance notice time, but the machine learning models (specifically the \ac{RDF} and \ac{LSTM}) still demonstrate significant predictive power even as far away as five seconds prior to the potential collision.

\subsection{Results for Research Question 3}\label{subsec:RQ3}
In Section~\ref{subsubsec:rudimentary_variant} we discuss the motivation for adding rudimentary paths to the dataset; namely, the desire to add realistic non-collision paths which would represent lane changes under naturalistic driving conditions.  In order to evaluate the efficacy of adding the rudimentary paths, we train a classifier using only the \textsc{vanilla}/\textsc{perturbed} hybrid-paths of the adversary and then test the classifier on the \textsc{rudimentary} hybrid-paths.  We expect some \textsc{rudimentary} paths (all of which are 0-labeled) to be misclassified due to their absence from the training set.  We find that only one \textsc{rudimentary} path was misclassified out of a total of 520 rudimentary paths in the testing set using a \ac{RDF}.  This indicates that perhaps \textsc{rudimentary} hybrid-paths don't significantly improve the performance and generalizability of the resulting models.

\subsection{Results for Research Question 4}\label{subsec:RQ4}
In Section~\ref{subsubsec:rudimentary_variant}, we discuss the motivation for adding \textsc{vanilla}/\textsc{perturbed} hybrid-path pairs of an adversary which are farther from the decision boundary.  We call these \textsc{variant} hybrid-path pairs, and one such pair is highlighted in Fig.~\ref{fig:decision_boundary}.  This allows the classifier to learn a broader notion of \textsc{vanilla} and \textsc{perturbed} adversary paths.  In order to evaluate the efficacy of adding the \textsc{variant} paths, we train a classifier using only the \textsc{vanilla}/\textsc{perturbed} adversary paths and then test the classifier on the \textsc{variant} paths.  We expect some of the \textsc{vanilla variant} and \textsc{perturbed variant} paths to be misclassified due to their distance from the existing paths present in the training set.  In a test set containing all \textsc{variant} paths, we find that the classifier still achieves an F1 score of 0.97 using the \ac{RDF} which was surprising.  The false negative rate did increase in this experiment from 0.01 to 0.05, indicating that adding the \textsc{variant} paths may better help the classifier identify collision paths that may be incorrectly classified as non-collision paths otherwise - this reduction in false negatives is a significant benefit for safety-focused \acp{CP}.  The results are shown in Table~\ref{table:ML_variants}.

\begin{table}[H]
    \vspace{0mm} 
    \caption[]{Results for Testing on Variant Paths Only}
    \centering
    \begin{tabular}{lccc} 
    \toprule 
    \textbf{Classifier} & \textbf{Recall} & \textbf{Precision} & \textbf{F1 Score} \\ \midrule
    RDF & 0.95 & 0.99 & 0.97 \\
    \bottomrule 
    \end{tabular}%
    \label{table:ML_variants}
\end{table}

\section{Discussion of Results}\label{sec:discussion_of_results}
Generally, this technique demonstrates that \acp{MLCP} can be utilized to assert about behaviors of \acp{CPS} in complex multi-agent systems such as self-driving vehicles, and these \acp{CP} can additionally be used to provide advance notice of critical events in order to mitigate hazards associated with failures of \acp{CPS}.

The question becomes, ``what then?''  Now that we have these \acp{MLCP}, what do we do with them?  There are two types of \ac{RM} which could be conducted to monitor these \acp{MLCP}: offline and online.

\subsection{Offline Runtime Monitoring}

Offline \ac{RM} is conducted retroactively, usually on log files recorded from the \ac{SUT}.  One of the most significant challenges to the widespread deployment of \acp{AV} is establishing liability for crashes and setting insurance rates.  For human drivers, this is typically done by looking at a driver's pattern of behavior to determine whether there is a high risk (and therefore should demand higher insurance rates), or whether it represents a lower risk of collisions.  When a collision does occur, the insurance company uses the data available to establish financial liability.  \Acp{CP} could play a key role in establishing risk metrics for \acp{AV} and assigning liability for collisions.  For example, suppose an insurer wants to determine the risk-taking level of an \ac{AV}; the insurer could train a \ac{MLCP} on naturalistic driving data that contains both normal and risky human driving behavior, and then use log files from \acp{AV} to determine how often the \acp{AV} exhibits risky driving behavior to better assess appropriate insurance rates.  

Similarly, suppose an \ac{AV} swerves suddenly on a highway to avoid a vehicle encroaching in its lane, and strikes a guardrail.  The insurer could run the \ac{MLCP} discussed in Section~\ref{subsec:RQ1} above to show that one second before the \ac{AV} swerved, a collision was imminent and the vehicle encroaching in the \ac{AV}'s lane was at fault and should bear the financial liability, even though it suffered no damage itself.  We do not attempt to weigh in on the discussion of \textit{who} exactly is liable for a fully autonomous system (e.g., the manufacturer, the owner), but many such \acp{MLCP} could be run offline for retrospective analysis to provide key risk metrics and behavior analysis to insurers, policymakers, vehicle manufacturers, regulators, and other stakeholders.

\subsection{Online Runtime Monitoring}\label{subsec:online}

Offline \ac{RM} is often the preferred approach for safety-critical systems because the computational power required to analyze the \acp{CP} in real time can interfere with the operation of the underlying system.  However, advances in edge computing, networking, and storage have mitigated this issue in many types of systems including \acp{AV}.  Online \ac{RM} can enable systems to respond to \ac{CP} violations in real time which in turn can aid in advancing cooperative autonomy, also known as human-machine teaming.  

As autonomous systems continue to advance and proliferate, humans who interact with such systems need to understand the tasks that can be delegated to them, along with the risks of delegating such tasks.  When humans place too much trust in the dependability of an autonomous system to perform a task, the human is not likely to maintain an adequate level of situational awareness in order to be able to intervene and take control such as in an emergency.  When humans place too little trust in autonomous systems, the cognitive burden (e.g., task overload) on humans is increased and the intended benefits of autonomy are reduced.  Therefore, understanding what tasks can be delegated to an autonomous system, the risks of delegating such tasks, and how that risk changes over time is critical to enabling humans to effectively trust autonomous systems.

This paradigm is well-illustrated by the current state of the automotive industry.  The \ac{SAE} defines six levels of driving automation ranging from Level 0 (no automation) up to  Level 5 (full automation) \cite{sae_j3016}.  Nearly all new vehicles in 2023 can be categorized as Level 2, in which the vehicle can perform steering and acceleration, but the human is solely responsible for monitoring the environment and must stay consistently engaged in the vehicle's control.  However, while the transition from Level 0 through 2 has proceeded relatively smoothly, and has even been welcomed by users and policymakers alike, the potential transition from Level 2 to Level 3 has been widely contested and the subject of much debate.  This is primarily because in Level 2, the human monitors the driving environment and is well-equipped to take over at any time.  In Level 3, the autonomous system monitors the driving environment, and the challenge of alerting a human concerning the need to take over control, provide awareness of the current state of the environment, and detect the need to do so far enough ahead of time, is a significant challenge.  Each of these factors: when to alert a human driver, whether or not the driver is professionally trained as an \ac{AV} backup driver, what to alert them about, how many alerts to provide, conditions under which they should be alerted, and time between alerts and takeover are all factors that can (and should) be encoded in \acp{CP}.

\Acp{MLCP},  defined in Section~\ref{subsec:terms} and described in Section~\ref{sec:results_analysis} can be well-suited for this purpose with additional human factors analysis.  Suppose an \ac{AV} manufacturer who is deploying Level 3 vehicles wants to warn a backup driver when a nearby vehicle is making an unsafe lane change.  In that case, the \ac{AV} would need to first detect the fact that a nearby vehicle is making an unsafe lane change, and then it would need to alert the backup driver through visual or haptic feedback in enough time for the driver to safely gain situational awareness and take control to avoid a potential collision.  Fig.~\ref{fig:ttc_vs_f1} demonstrates that a classifier still performs well when attempting to predict whether or not there will be a collision even up to five seconds away from the potential collision, demonstrating that \acp{MLCP} could be suitable for triggering real time modifications to the behavior of such \acp{CPS}, including warning humans if present.

\section{Future Work}\label{sec:future_work}

In the experiments discussed in Section~\ref{sec:experiment}, we exclude uncertainty related to sensor readings as a confounding factor.  Therefore, the features which represent the kinematics of the adversary vehicles are collected directly from the simulation, and are not the result of sensor readings taken by the ego vehicle.  Such an approach is well suited for much-discussed trend of ``connected highways'' and machine-to-machine/vehicle-to-vehicle communication \cite{hameed_2017}.  It also makes way for future work where the \textit{HybridPair} search is conducted using state-of-the-art sensors.  

Additionally, the detection window is defined as a certain time prior to the potential collision, but in real-world operation, it is unclear when a potential collision might occur, making it unclear when the detection window begins for evaluating the \ac{MLCP}.  Therefore, it would be necessary to define preconditions for the \ac{MLCP} to begin its prediction window, which could be triggered by the proximity of a nearby vehicle to the ego.

Various times between a \ac{CP} flagging and a potential collision are discussed.  There is significant room for research into the time required between alerting a human of the need to take control and actually transferring control from an autonomous system to a human.  In addition, the \acp{MLCP} could be trained to vary their detection window depending on the situational awareness of the human backup driving as measured by an engagement metric such as eye tracking.

Finally, the results for Research Question 3 (Section~\ref{subsec:RQ3}) could indicate that despite our efforts in create a high-variance training set, the training set still needs additional factors to provide sufficient variance.  Such factors could be related to the addition of weather, varying road friction coefficients, lighting conditions, junction types, non-vehicle actors, etc.

\section{Conclusion}\label{sec:conclusion}
While once only developed by deep-pocket corporations and government agencies like the  \ac{NASA}, software-based \acp{CPS} are now ubiquitous.  Autonomous \acp{CPS} are being deployed at a rapid pace; the automotive industry estimates that over 740,000 autonomous-ready vehicles will be on global roads by the end of 2023, and in early 2023.  Providing continuous assurance of autonomous systems is of great interest to the military as well.  In early 2023, the \acs{US} \ac{DOD} released an updated strategy for deploying lethal weapons systems with autonomous capabilities with a requirement that such systems undergo rigorous hardware and software \ac{VV} \cite{DODI_3000_09}.  Explainable \acp{MLCP} provide a way to make it practical to apply formal methods on real-world  \acp{CPS} and intelligent transportation systems, providing continuous assurance of dependable behaviors throughout the entire system lifecycle.

\section*{Disclaimer}
The views and conclusions contained herein are those of the authors and should not be interpreted as necessarily representing the official policies or endorsements, either expressed or implied, of the U.S. Government. The U.S. Government is authorized to reproduce and distribute reprints for Government purposes notwithstanding any copyright annotations thereon.

\bibliographystyle{IEEEtran}
\bibliography{IEEEabrv,references}

\end{document}